\documentclass[letter,        
               keeplastbox,   
               ]{jacow}
\usepackage{pdfpages,multirow,ragged2e} %
\makeatletter%
	\ifboolexpr{bool{xetex}}
	 {\renewcommand{\Gin@extensions}{.pdf,%
	                    .png,.jpg,.bmp,.pict,.tif,.psd,.mac,.sga,.tga,.gif,%
	                    .eps,.ps,%
	                    }}{}
\makeatother

\ifboolexpr{bool{xetex} or bool{luatex}} 
 {}                                      
 {\usepackage[utf8]{inputenc}}           

\usepackage[USenglish]{babel}

\makeatletter
\@namedef{ver@lineno.sty}{9999/12/31}
\@namedef{opt@lineno.sty}{}
\makeatother
\usepackage[frozencache=true,cachedir=minted-cache]{minted}  
\usepackage{float}   

\ifboolexpr{bool{jacowbiblatex}}%
 {%
  \addbibresource{jacow-test.bib}
  \addbibresource{biblatex-examples.bib}
 }{}
\begin{document}

\title{Next Generation Computational Tools for the Modeling and Design of Particle Accelerators at Exascale\thanks{Work supported by the Exascale Computing Project (17-SC-20-SC), a joint project of the U.S. Department of Energy's Office of Science and National Nuclear Security Administration, responsible for delivering a capable exascale ecosystem, including software, applications, and hardware technology, to support the nation's exascale computing imperative. This work was supported by the Laboratory Directed Research and Development Program of Lawrence Berkeley National Laboratory under U.S. Department of Energy Contract No. DE-AC02-05CH11231.
This research used resources of the National Energy Research Scientific Computing Center (NERSC), a U.S. Department of Energy Office of Science User Facility located at Lawrence Berkeley National Laboratory, operated under Contract No. DE-AC02-05CH11231 using NERSC award HEP-ERCAP0020744.
}}

\author{A. Huebl\thanks{axelhuebl@lbl.gov}, R. Lehe, C. E. Mitchell, J. Qiang, R. D. Ryne, R. T. Sandberg and J.-L. Vay\\
Lawrence Berkeley National Laboratory, 94720 Berkeley CA, USA}

\maketitle

\begin{abstract}
   Particle accelerators are among the largest, most complex devices.
To meet the challenges of increasing energy, intensity, accuracy, compactness, complexity and efficiency, increasingly sophisticated computational tools are required for their design and optimization.
It is key that contemporary software take advantage of the latest advances in computer hardware and scientific software engineering practices, delivering speed, reproducibility and feature composability for the aforementioned challenges.
A new open source software stack is being developed at the heart of the Beam pLasma Accelerator Simulation Toolkit (\verb+BLAST+) by LBNL and collaborators, providing new particle-in-cell modeling codes capable of exploiting the power of GPUs on Exascale supercomputers.
Combined with advanced numerical techniques, such as mesh-refinement, and intrinsic support for machine learning, these codes are primed to provide ultrafast to ultraprecise modeling for future accelerator design and operations.
\end{abstract}

%

\section{Introduction}
Large-scale computer simulations of charged particle motion inside of particle accelerators play a crucial role in accelerator design and operation.
In order to quickly simulate charged particle dynamics, including collective effects, advanced software must be developed to take advantage of state-of-the-art computer hardware.

With the onset of the Exascale supercomputing era, omnipresent GPU-accelerated machines require multi-level parallelism, multi-paradigm programming and dynamic load balancing.
The U.S. Department of Energy Exascale Computing project addressed this need by co-developing applications and software for Exascale computing acquisitions.
The laser-plasma modeling code \verb+WarpX+~\cite{warpx}, e.g., used in plasma-based particle acceleration, is a result of this project and recently provided the first full-scale runs at the first demonstrated Exascale machine, Frontier~\cite{gb22}.

\subsection{Beam, Plasma \& Accelerator Simulation Toolkit}

\verb+WarpX+ is a code in the Beam, Plasma \& Accelerator Simulation Toolkit (\verb+BLAST+, \url{https://blast.lbl.gov}), a suite of open source particle accelerator modeling codes.
Originally developed under the name Berkeley Lab Accelerator Simulation Toolkit, included codes achieved compatibility through a common meta-data standard in I/O, \verb+openPMD+\cite{openPMD}, yet were implemented in disjoint code bases.
\verb+BLAST+ has been renamed in 2021 to reflect international contributions from LIDYL (CEA, France), SLAC (USA), LLNL (USA), DESY (Germany), UHH (Germany), HZDR (Germany), Radiasoft (USA), CERN (Switzerland) and more; \verb+BLAST+ development involves deep collaboration among physicists, applied mathematicians, and computer scientists.

With the emergence of the first Exascale Computing supercomputers, modeling codes that were originally designed for parallel CPU-powered machines need to undergo a fundamental modernization effort.
This became necessary, as compute nodes are now equipped with accelerator hardware such as GPUs (and potentially FPGAs in the future).
Selected as application for the Department of Energy Exascale Computing Project, the \verb+BLAST+ code \verb+WARP+\cite{warp1,warp2} underwent a complete rewrite from Fortran to modern C++ resulting in its successor \verb+WarpX+\cite{warpx}.
Building on the momentum of this transition to form a more cohesive Accelerator Toolkit, the specialized plasma wakefield acceleration code \verb|HiPACE++|\cite{hipacepp} and beam dynamics code \verb+ImpactX+\cite{impactxrepo}, presented herein, are developed.

\subsection{Software Design}

\begin{figure}[ht]
\centering
    \vspace{-11pt}
\includegraphics[width=\linewidth]{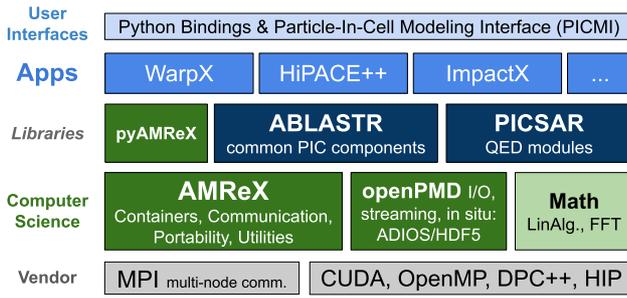}
    \vspace*{-11pt}
    \caption{\small Design of the \texttt{BLAST} software stack.
  Modularization enables code sharing and tight coupling.}
\label{fig:blast}
\end{figure}

A central goal of the modernization of \verb+BLAST+ is modularity for efficient code reuse and tight integration for coupling, i.e., in hybrid particle accelerators with conventional and advanced (plasma) elements.
Figure~\ref{fig:blast} shows the design of \verb+BLAST+'s software dependencies, with upper components depending and sharing lower blocks in the schema.
Shared code, common application programming interfaces (APIs) and data standards ensure composability and connection to the AI/ML and data science ecosystems.
Performance-critical routines are implemented and reused in modern C++, using a single-source approach to program both CPUs and GPUs via a performance-portability layer in \verb+AMReX+~\cite{amrex}.
The newly introduced \verb+ABLASTR+ library collects common particle-in-cell (PIC) routines.

Python high-level interfaces are used for user efficiency and to provide standardized APIs to data science and AI/ML frameworks, which are mostly driven from the same language.
Documentation and examples are developed in lock-step with documentation and published on \url{https://impactx.readthedocs.io}.
Examples and test cases are continuously run against expected results.

All development is carried out in the open using open source licenses, contributable code repositories, code reviews, regular releases and change logs~\cite{impactxrepo}.
The community reports open ``issues'' for feature requests, bug reports, etc.
Installation for users and developers is supported by package managers and HPC modules.

\section{ImpactX}

\verb+ImpactX+ is an $s$-based beam dynamics code.
Leveraging expertise and models in \verb+IMPACT-Z+~\cite{IMPACTZ} and \verb+MaryLie+~\cite{MaryLie, MLI}, this new simulation code is built from the ground up to take GPU-accelerated computing, mesh-refinement for space-charge effects, load balancing, and coupling AI/ML frameworks into consideration.
\verb+ImpactX+ design relies on open community standards for I/O and data interfaces.

As of version 22.08, many features are still under active development.
Generalized and reused from \verb+WarpX+ via the \verb+ABLASTR+ library are GPU-accelerated routines for charge deposition, beam statistics, Poisson solve, profiling, warning logging, Unix signal handling, build/installation logic, among others.

\subsection{Model Assumptions}

Tracking through lattice elements is performed by pushing particles in $s$ using a symplectic map.
Currently, each map applied during tracking is accurate through linear order (with respect to the reference orbit).
The linearization implies that, when solving for space-charge effects, we assume that the relative spread of velocities of particles in the beam is negligible compared to the velocity of the reference particle.

The space charge fields are treated as electrostatic in the bunch rest frame.
In particular, no retardation or radiation effects are included, and we solve the Poisson equation in the bunch rest frame.
The effect of space charge is included in tracking using a map-based split-operator approach~\cite{IMPACTZ}.

\subsection{Usage Example}

\verb+ImpactX+ can be executed in two ways: a traditional executable reading a textual input file or driven from Python.
The latter approach is compact and expressive for %
%
\begin{listing}[H]
\begin{minted}[linenos,
               numbersep=5pt,
			   %fontsize=\scriptsize,
               frame=lines,
               framesep=2mm]{python3}
from impactx import ImpactX, RefPart, \
                    distribution, elements

sim = ImpactX()  # simulation object

# set numerical parameters and IO control
sim.set_particle_shape(2)  # B-spline order
sim.set_slice_step_diagnostics(True)
sim.set_space_charge(False)

# domain decomposition & space charge mesh
sim.init_grids()

# load a 2 GeV electron beam with an initial
# unnormalized rms emittance of 2 nm
energy_MeV = 2.0e3  # reference energy
charge_C = 1.0e-9  # used with space charge
mass_MeV = 0.510998950  # mass
qm_qeeV = -1.0e-6/mass_MeV  # charge/mass
npart = 10000  # number of macro particles

distr = distribution.Waterbag(
    sigmaX = 3.9984884770e-5,
    sigmaY = 3.9984884770e-5,
    sigmaT = 1.0e-3,
    sigmaPx = 2.6623538760e-5,
    sigmaPy = 2.6623538760e-5,
    sigmaPt = 2.0e-3,
    muxpx = -0.846574929020762,
    muypy = 0.846574929020762,
    mutpt = 0.0)
sim.add_particles(
    qm_qeeV, charge_C, distr, npart)

# set the energy in the reference particle
sim.particle_container().ref_particle() \
    .set_energy_MeV(energy_MeV, mass_MeV)

# design the accelerator lattice
ns = 25  # steps slicing through ds
fodo = [
    elements.Drift(ds=0.25, nslice=ns),
    elements.Quad(ds=1.0, k=1.0, nslice=ns),
    elements.Drift(ds=0.5, nslice=ns),
    elements.Quad(ds=1.0, k=-1.0, nslice=ns),
    elements.Drift(ds=0.25, nslice=ns)
]
# assign a fodo segment
sim.lattice.extend(fodo)

# run simulation
sim.evolve()
\end{minted}
  \captionof{listing}{
An example Python script \texttt{FODO.py} that can be used to design an FODO cell setup with \texttt{ImpactX} v22.08.
This script can equally drive CPU and GPU simulations and executes over multiple nodes if started with MPI.}
  \label{lst:impl}
\end{listing}
simulation design, since Python's well-known syntax can be used to design complex beamlines with lines and segments.
From a performance viewpoint, both are equivalent since computational C++ kernels are compiled and just wrapped from Python.
Listing~\ref{lst:impl} shows such a script-driven simulation.

Furthermore, such scripts can be extended since \verb+ImpactX+ exposes the underlying \verb+AMReX+ data structures to Python via \verb+pyAMReX+, implementing AMReX data storage for zero-copy access via a standardized \textit{Array Interface}~\cite{arrayinterface}.
For instance, beam particles and generated fields can be manipulated and analyzed in memory, combined with solvers from other \verb+BLAST+ or third party software packages and additional computational routines can be added, even with GPU support, using packages such as \verb+cupy+ or \verb+numba+.

\section{Numerical Experiments}


Correctness tests are performed continuously, for every code change, on \verb+ImpactX+.
The goal of these tests is to cover all implemented functionality and verify that computed results are within expected precision, independently of the compute hardware used.
Following such \textit{test-driven development} eases the entry burden for accelerator scientists adding new functionality to the project, since automated testing will inform them if unexpected side-effects of changed code would change benchmarked physics results.
Tests and examples also add a solid body of documented examples to the project.

As of version 22.08, the following benchmarks and examples are implemented: a FODO cell, a magnetic bunch compression chicane~\cite{Chicane}, a stationary beam in a constant focusing channel, a Kurth-distribution beam in a periodic isotropic focusing channel~\cite{Kurth}, a stable FODO cell with short RF~(buncher) cavities added for longitudinal focusing~\cite{RyneBenchmarks}, a chain of thin multipoles, a nonlinear focusing channel based on the IOTA nonlinear lens, and a model of the Fermilab IOTA storage ring~(linear optics)~\cite{IOTA}.
Detailed numerical parameters are archived online~\cite{impactxrepo,dataartifact}.

\subsection{Benchmark: FODO Cell}

\begin{figure}[H]
  \centering
  \includegraphics[width=\columnwidth]{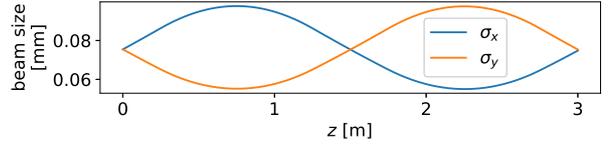}
  \caption{Evolution of the horizontal and vertical rms beam sizes in the FODO benchmark.
  \label{fig:fodo-sigma}}
\end{figure}

\begin{figure}[H]
  \centering
  \includegraphics[width=\columnwidth]{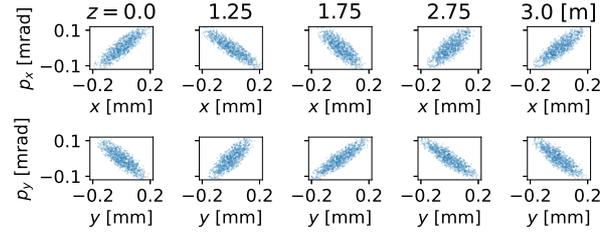}
  \caption{Transverse phase space projections in the FODO cell benchmark.
  \label{fig:fodo-scatter}}
\end{figure}

In this benchmark, a stable FODO lattice with a zero-current phase advance of 67.8 degrees per cell is modeled.
An rms-matched \SI{2}{GeV} electron beam with initial unnormalized rms emittance of \SI{2}{nm} propagates through a single cell, with the beam size evolution as shown in Fig.~\ref{fig:fodo-sigma}.
In Fig.~\ref{fig:fodo-scatter}, the evolution of the transverse phase space is shown.
The benchmark verifies for every code change that the beam second moments remain matched and that the emittances remain constant.


\subsection{Benchmark: Berlin-Zeuthen Chicane}

\begin{figure}[H]
  \centering
  \includegraphics[width=\columnwidth]{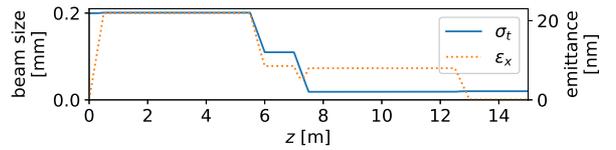}
  \caption{Evolution of longitudinal beam size~(rms) and transverse emittance in the chicane benchmark.
  \label{fig:chicane-sigma}}
\end{figure}

\begin{figure}[H]
  \centering
  \includegraphics[width=\columnwidth]{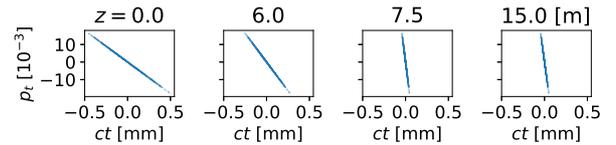}
  \caption{Evolution of the longitudinal phase space in the chicane benchmark, illustrating $10\times$ compression.
  \label{fig:chicane-scatter}}
\end{figure}

This benchmark is a simple, unshielded four-bend chicane with parameters similar to the ones required for the compression stages at LCLS~(at \SI{5}{GeV}) or TESLA XFEL~(at \SI{500}{MeV})~\cite{Chicane}.
Here, a \SI{5}{GeV} electron bunch with normalized transverse rms emittance of \SI{1}{\um} is used.
Figure~\ref{fig:chicane-sigma} shows the longitudinal beam size and transverse emittance evolution.  The former is compressed $10\times$, while the initial emittance is recovered at the end of transport.
The longitudinal phase space evolution during the transport is shown in Fig.~\ref{fig:chicane-scatter}.


\subsection{Benchmark: Fermilab IOTA Storage Ring}

This benchmark is a model of the bare~(linear) lattice of the Fermilab IOTA storage ring~(v. 8.4)~\cite{IOTA}, with optics configured for operation with a \SI{2.5}{MeV} proton beam.  An rms-matched proton beam with an unnormalized emittance of \SI{4.5}{\micro m} propagates over a single turn.
The second moments of the particle distribution after a single turn are checked to coincide with the initial second moments of the particle distribution, to within the level expected due to numerical particle noise.

\begin{figure}[H]
  \centering
  \includegraphics[width=\columnwidth]{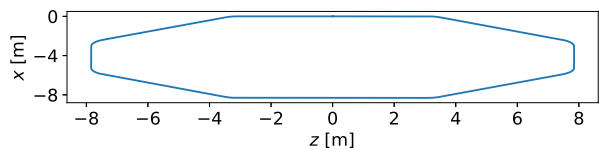}
  \caption{The reference orbit in the IOTA lattice benchmark, as produced by \texttt{ImpactX}, showing the storage ring floorplan.
  \label{fig:iota-position}}
\end{figure}

\begin{figure}[H]
  \centering
  \includegraphics[width=\columnwidth]{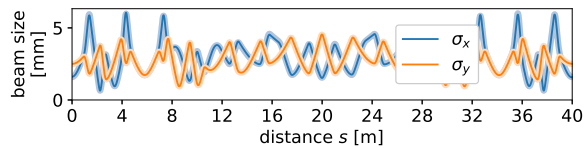}
  \caption{Evolution of the rms beam size in the IOTA lattice benchmark.
  Thin dark lines: \texttt{ImpactX}, light bold lines: \texttt{IMPACT-Z} results.
  \label{fig:iota-sigma}}
\end{figure}

In Fig.~\ref{fig:iota-position}, the reference orbit indicating the global beam position within the ring is shown.
Figure~\ref{fig:iota-sigma} shows the rms beam size evolution as a function of path length over a single turn.
The thin dark lines are from \verb+ImpactX+, while the light bold lines in the background are from \verb+IMPACT-Z+. The results of the two codes are in excellent agreement.


\subsection{Performance}

Although \verb+ImpactX+ in version 22.08 is in an early development state, initial performance comparisons can be made between conventional~(CPU) and accelerated compute hardware~(GPU).
In the following, the above IOTA lattice benchmark is used as a computing benchmark~(without I/O or space charge solvers) to measure the performance of \verb+ImpactX+ on a node of the NERSC Perlmutter Phase 2 supercomputer.
Prototypical for future large-scale simulations, but not strictly needed for this benchmark to converge, $10^8$ beam macro particles are used.

\begin{figure}[H]
  \centering
  \includegraphics[width=\columnwidth]{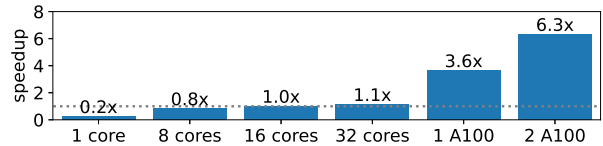}
  \caption{Relative performance between CPU and GPU runs on Perlmutter Phase 2~(NERSC).
  16 CPU cores~(w/o hyperthreading) and one A100 GPU are each 1/4th of a node's resources: one AMD EPYC 7763~(Milan) processor and four NVIDIA Ampere A100 SXM2~(\SI{40}{GB}) GPUs.
  \label{fig:iota-performance}}
\end{figure}

Figure~\ref{fig:iota-performance} shows the \verb+ImpactX+ strong-scaling speedup in time-to-solution relative to a 16 CPU core run~(146\,sec).
Compilation uses \verb+-O3+ and fast-math enabled; compilers are GNU 11.2.0 and NVCC 11.7.64, respectively.
Values above $1.0$ are faster, below are slower.
Dynamic load-balancing is not yet used.

\subsection{Data Availability}

Simulation code and documentation are openly developed in Ref.~\cite{impactxrepo}.
Numerical experiments and performance results are archived in Ref.~\cite{dataartifact}.

\section{CONCLUSION}

This paper presents computational tools for the modeling and design of particle accelerators, readying codes up for next generation machines in the Exascale era.
The open source software toolkit \verb+BLAST+ provides modeling tools to model hybrid accelerators, containing both plasma and conventional beamline elements.
\verb+ABLASTR+ is a modern C++17 library used to share particle-in-cell routines between simulation codes.
Based on this, \verb+ImpactX+ is developed to succeed \verb+IMPACT-Z+ as a new, $s$-based beam dynamics code with intrinsic GPU, mesh-refinement and tight coupling to time-based codes and AI/ML capabilities.

\verb+ImpactX+ is in an early state, but already able to model significantly larger particle ensembles than its predecessor codes.
Ongoing and future developments will add capabilities for: space charge effects, scalable I/O, non-linear elements, RF-modeling, wakefield effects, surface methods and support to read accelerator lattices from MAD-X files, among others.

\section{ACKNOWLEDGEMENTS}
We acknowledge the community of contributors to the open source projects \verb+ABLASTR+, \verb+AMReX+, \verb+ImpactX+, \verb+IMPACT-Z+, \verb+openPMD+, \verb+pybind11+ and \verb+WarpX+.

%
%
\ifboolexpr{bool{jacowbiblatex}}%
	{\printbibliography}%
	{%

} 

\end{document}